\documentclass[aps,prl,epsfig,floats,amssymb,amsmath,twocolumn,showpacs,footinbib]{revtex4-1}

\begin{document}

\title{Reply to Comment on ``Enhanced diffusion of enzymes that catalyze exothermic reactions''}

\author{Ramin Golestanian}
\affiliation{Rudolf Peierls Centre for Theoretical Physics, University of Oxford, 1 Keble Road, Oxford, OX1 3NP, UK}

\begin{abstract}
Catalytically active enzymes have recently been observed to exhibit enhanced diffusion. In a recent work [C. Riedel et al., Nature {\bf 517}, 227 (2015)], it has been suggested that this phenomenon is correlated with the degree of exothermicity of the reaction, and a mechanism was proposed to explain the phenomenon based on channeling the released heat into the center of mass kinetic energy of the enzyme. I addressed this question by comparing four different mechanisms, and concluded that collective heating is the strongest candidate out of those four to explain the phenomenon, and in particular, several orders of magnitude stronger than the mechanism proposed by Riedel et al. In a recent preprint (arXiv:1608.05433), K. Tsekouras, C. Riedel, R. Gabizon, S. Marqusee, S. Press\'e, and C. Bustamante present a comment on my paper [R. Golestanian, Phys. Rev. Lett. {\bf 115}, 108102 (2015); arXiv:1508.03219], which I address here in this reply.
\end{abstract}
\date{\today}


\maketitle

In a recent paper by Riedel et al. \cite{Bustamante}, empirical evidence was presented to suggest that the recently discovered phenomenon of enhanced diffusion of catalytically active enzymes correlates with the amount of heat of reaction released into the solution during the catalysis, and a theoretical explanation was proposed for the phenomenon. By examining this phenomenon from a number of different angles, I presented a critique of their theoretical proposal in Ref. \cite{RG} and argued that a systematic derivation of their result yields values for the diffusion enhancement that are 6-7 orders of magnitude too small. Moreover, I suggested that two other mechanisms, namely stochastic swimming and collective heating, are stronger than the proposed mechanism by Riedel et al. \cite{Bustamante}. In a comment \cite{comment}, Tsekouras et al state that they have eliminated alternative explanations of their results through experimental
controls, implying that they believe their acoustic wave mechanism is the strongest candidate. I disagree with this statement, which involves numerical evaluations of various contributions that have not been performed consistently for all mechanisms (as discussed in Ref. \cite{RG}), favoring one specific mechanism over the others without sufficient evidence. Moreover, this assertion is based on the assumption that there is only one mechanism that governs all the observed experiments, whereas the complexity of the system suggests that there are likely several contributions at work which could add up with varying degrees of significance depending on the particular condition in each experiment. 

%\paragraph{Transient versus stationary regime of collective heating.}
Tsekouras et al  \cite{comment} bring up the point that in the case of collective heating, I have done my estimate for a stationary state while their experiment is not yet in stationary state; it is in a transient state. In Ref. \cite{RG}, I have briefly discussed what happens when one needs to consider the transient behavior of the resulting nonlinear heat diffusion equation, using an analogy to the phenomenology of flame propagation. The equation admits a propagating wave solution, reminiscent of the Fisher-PKK equation. This occurs as a quick nucleation of a stationary solution that is separated by a sharp boundary from a null solution, with the front moving at a characteristic velocity that is defined as the square root of the product of the effective rate and the heat diffusion coefficient. Using the numbers I use in Ref. \cite{RG}, this speed comes out as $\sim 1 \, \mu{\rm m}/s$. Therefore, the prediction will be that when the effect is observed, it is presumably because the illuminated spot in which diffusion is probed is inside the stationary part of the wave front, and thus for that part the stationary solution will suffice as an estimate. The speed is fast enough to justify rapid observation of the enhancement effect in the observation domain. The transient nature will then also mean that this effect will decay with substrate depletion. This effect can simply be incorporated in the estimation by taking into account the effective catalytic reaction rates over the period of observation.

%\paragraph{Has collective heating been experimentally refuted?}
The control experiment performed by Riedel et al  \cite{Bustamante} does not rule out collective heating. Unlike the narrative of the observation in Ref. \cite{Bustamante}, the experiments on unlabeled active catalase plus labeled inactive urease do seem to show an increase in the effective diffusion coefficient consistent with a few percentage rise that is typical in these systems, albeit with large error bars. It is not clear why they have chosen to ignore this clear trend in Ref. \cite{Bustamante}. Interestingly, the original paper by Ayusman Sen's group \cite{Sen-2} shows a similar control experiment that indeed has an exact same trend of increase by a few percent for ``tracers''; see Fig. 8B in Ref. \cite{Sen-2}. Considering that the other mechanisms (including the proposed mechanism discussed in Ref. \cite{Bustamante}) are several orders of magnitude off scale when one evaluates the quantities by realistic numbers, this suggestive trend that exhibits the right order of magnitude cannot be ignored. 

Tsekouras et al  \cite{comment} mention the observation by Sen et al of separation of active enzymes \cite{Sen-1} as proof that collective heating cannot explain enhanced diffusion. I have a number of comments on this statement. First, to my knowledge there is as yet no theoretical explanation of the intriguing observation of active enzyme separation. Second, it is not a priori clear that this phenomenon is governed by the same mechanism that leads to enhanced diffusion. Third, my understanding is that the acoustic wave mechanism proposed in Ref. \cite{Bustamante} cannot explain the separation phenomenon, so I do not see how this point is relevant to the current debate. The existence of this unexplained phenomenon, however, does indicate that there are more things to discover about these fascinating systems, and it is likely that as yet unknown mechanisms will close the gaps in all these discussions.

It is helpful to re-iterate the message that I had intended for Ref. \cite{RG} to convey. In that paper, I have identified four possible mechanisms that can explain the observation (there could certainly be others), quantified their contributions using simple estimates that could certainly be improved upon by adding more realistic details, and ordered them in terms of magnitude. The conclusion is that heating comes out as the strongest, and the mechanism proposed by Riedel et al. \cite{Bustamante} based on acoustic waves comes out as the third in line, and 6-7 orders of magnitude too small if we take into account energy partitioning between deformation modes. While taking into account more realistic aspects such as the specific boundary condition and the transient nature of the experiment etc could change these estimates by an order of magnitude, I do not expect this ordering to change when one makes more or less conservative estimates, because of the separation of the orders of magnitude between them. In Ref. \cite{Bustamante}, a choice has been made to favor the acoustic wave mechanism over heating, while the former is five orders of magnitude smaller than the latter.

\end{document}